\documentclass[10pt,aps,prl,twocolumn,superscriptaddress,floatfix,longbibliography]{revtex4-1}

\usepackage{graphicx}
\usepackage{color}
\usepackage{amsmath}
\usepackage{amssymb}
\usepackage{amsfonts}
\usepackage{comment}
\usepackage{ulem}
\usepackage{braket}
\definecolor{scarred}{rgb}{0.75,0.0,0.0}
\usepackage{hyperref}
\hypersetup{
colorlinks=true,final=true,
        linkcolor=blue,
        citecolor=blue,
        filecolor=blue,
        urlcolor=blue,
}

\usepackage{mathtools}
\usepackage{amsbsy}

\begin{document}

\title{Co-operating multiorbital and nonlocal correlations in bilayer nickelate} 

\author{Evgeny A. Stepanov}
%\email{evgeny.stepanov@polytechnique.edu}
\affiliation{CPHT, CNRS, \'Ecole polytechnique, Institut Polytechnique de Paris, 91120 Palaiseau, France}
\affiliation{Coll\`ege de France, 11 place Marcelin Berthelot, 75005 Paris, France}
\author{Steffen B\"otzel}
\affiliation{Theoretische Physik III, Ruhr-Universit\"at Bochum, D-44780 Bochum, Germany}
\author{Ilya M. Eremin}
\affiliation{Theoretische Physik III, Ruhr-Universit\"at Bochum, D-44780 Bochum, Germany}
  \author{Frank Lechermann}
\affiliation{Theoretische Physik III, Ruhr-Universit\"at Bochum,
  D-44780 Bochum, Germany}  

\begin{abstract}
The interplay of multiorbital physics and nonlocal self-energy effects is studied within an effective three-orbital model for the high-pressure normal state of superconducting bilayer nickelate La$_3$Ni$_2$O$_7$. The
model is solved within an advanced many-body framework capturing $k$-dependent correlations beyond dynamical mean-field theory. Different low-energy scenarios subtly depend on the strength of the interorbital interaction, either placing the notorious flat $\gamma$ quasiparticle band in the occupied part of the spectrum, or letting it cross the Fermi level. In the latter case, intriguing spin-polaron formation due to the scattering of electrons with paramagnon excitations takes place. This leads to bound states appearing as a shadow band with incoherent low-energy spectral weight below the Fermi level. Our results uncover additional competing states that exist in bilayer nickelates and could explain the controversy of recent angle-resolved photoemission experiments. \end{abstract}

\maketitle

%{\it Introduction.} 
Nonlocal electronic correlation effects are relevant features of various (quasi-)low-dimensional interacting lattice problems, including for example the quantum magnetism phenomenon in two spatial dimensions (2D)~\cite{sachdev08}. The very rich and intriguing phase diagram of high-Tc superconducting cuprates~\cite{keimer15}
is hard to grasp without an appreciation of electronic correlations beyond the local limit. For instance, within the Mott scenario of effective single-orbital
cuprates, near-neighbor correlations are essential to account for basic salient features, as e.g. provided by a cluster dynamical-mean field theory (DMFT) picture~\cite{lichtkat00,parcollet04}.
On the other hand, in manifest multiorbital systems,
such as high-T$_c$ iron-based superconductors~\cite{fernandes22}, nonlocal processes may more likely be overruled by local-orbital based correlation mechanisms.

\begin{figure}[t]
\includegraphics[width=\linewidth]{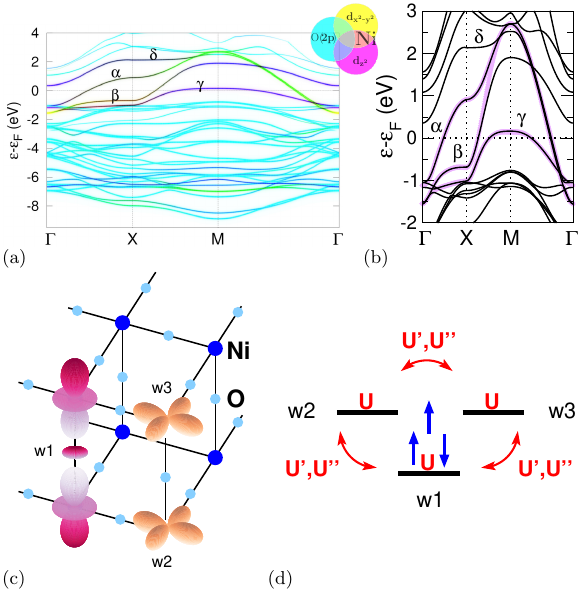}
\caption{Model setup for La$_3$Ni$_2$O$_7$. (a) DFT band structure in fatspec representation for Ni-$d_{z^2}$, Ni-$d_{x^2\text{-}y^2}$ and O$(2p)$-orbitals. (b) Low-energy DFT bands (black) with threefold Wannier dispersion (purple) according to $\{w1,w2,w3\}$ orbitals, schematically shown in (c). (d) Pictogram for the interacting part of the minimal three-orbital model, with degenerate $w2,w3$ levels and a deeper-lying $w1$ level. Note that $U'=U''+J$.}
\label{pfig1}
\end{figure}

In this regard, the recently discovered nickelate superconductor La$_3$Ni$_2$O$_7$ under high pressure~\cite{sun23} or as compressively-strained thin film~\cite{ko24, zhou-ambient24} poses an interesting problem. As revealed from density functional theory (DFT) [see Fig.~\ref{pfig1}\,(a)], its low-energy electronic structure builds up from three electrons distributed over Ni-$e_g\,\{d_{z^2},d_{x^2\text{-}y^2}\}$ hybridized with O$(2p)$, giving rise to two antibonding Ni-$d_{x^2\text{-}y^2}$-dominated 
$(\alpha,\beta)$ bands, one antibonding Ni-$d_{z^2}$-dominated $\delta$ band and one nonbonding Ni-$d_{z^2}$-dominated flat $\gamma$ band~\cite{lechermann2025,foyevtsova2025,devaulx2025,Hu2025}.
While multiorbital physics appears therefore inevitable for this formal Ni($3d^{7.5}$) bilayer compound~\cite{luohu23,lechermann23,zhanglin23,shilenko23,chen-jul23,sakakibara24,christiansson23,yangwangwang23,lupan23,liumei23,yangzhang23,xingzhou24,jianghuo23,luobiao23,oh24,ryee24,Savrasov2024,geisler24,bleys25}, the role of nonlocal correlations remains largely obscure. There are Ni-$d_{z^2}$ singlet-forming interlayer correlations as also revealed by cluster-DMFT~\cite{ryee24} and cluster-auxiliary-boson calculations~\cite{lechermann2025}, but robust singlet formation appears implausible due to significant $d_{z^2}$-$d_{x^2\text{-}y^2}$ hybridization. Beyond the short-range regime, tensor-network and dynamical matrix renormalization group studies~\cite{shen-dmrg23, xingzhou24} may so far only tackle reduced orbital models for La$_3$Ni$_2$O$_7$, and are moreover challenging for non-1D architectures.

In this work, we argue that the normal state of bilayer La$_3$Ni$_2$O$_7$ represents a remarkable case where the interplay of multiorbital and nonlocal correlations may lead to intriguing low-energy physics. First, the location of the $\gamma$ band in energy depends strongly on the interorbital interaction. Second, when crossing the Fermi level, the itinerant electrons scatter with ferromagnetic (FM) spin fluctuations originating from the flat-band character of this band, resulting in spin-polaron bound states~\cite{katsnelson82, PhysRevB.91.155114, FM_doping}. Those composite excitations show up as incoherent spectral weight below the Fermi level.

{\it Model and methods.}
The DFT band structure marks the $\alpha,\beta,\gamma$
bands as forming the threefold La$_3$Ni$_2$O$_7$ fermiology. As it was shown in Ref.~\cite{lechermann2025}, this three band dispersion may effectively described by the maximally-localized Wannier orbitals $w1$, $w2$, $w3$ [see Fig.~\ref{pfig1}\,(b,\,c)] in a minimal-cluster picture. While $w2$, $w3$ are Wannier variations of Ni-$d_{x^2\text{-}y^2}$ hybridized with O-$2p_{x,y}$ in
respective layers, the $w1$ orbital derives from the nonbonding
Ni-$d_{z^2}-$O-$p_z$$-$Ni-$d_{z^2}$ molecular orbital across the bilayer. Hence, the latter orbital dominantly pictures the $\gamma$ band. An effective three-orbital Hubbard Hamiltonian may thus be written as
\begin{align}
H = &\sum_{jj',\sigma,ll'} t^{ll'}_{jj'} c^{\dagger}_{jl\sigma} c^{\phantom{\dagger}}_{j'l'\sigma} 
+ \sum_{j,l} U n_{jl\uparrow} n_{jl\downarrow} \notag\\
&+ \frac12 \sum_{j,l\neq{}l'}\sum_{\sigma,\sigma'} (U''+J\delta_{\sigma,\sigma'}) n_{jl\sigma} n_{jl'\sigma'}, \notag
%\label{eq:H_latt}
\end{align}
where $c^{(\dagger)}_{jl\sigma}$ describes annihilation (creation) of an electron on the site $j$ on the Wannier orbital ${l\in\{w1,w2,w3\}}$ with spin projection ${\sigma\in\{\uparrow,\downarrow\}}$.
The hopping amplitudes $t^{ll'}_{jj'}$ are obtained from the Wannier downfolding and we focus on the freestanding bilayer, neglecting $k_z$ dispersions.
The interaction part is considered in the local density-density approximation, where $U$ is the intraorbital on-site Coulomb repulsion and ${U'' = U-3J}$ is the interorbital Coulomb repulsion on the minimal two-site cluster. The additional interaction term proportional to ${J\delta_{\sigma,\sigma'}}$ reflects the competition between the usual Hund coupling $J_{\rm H}$ and interlayer spin interactions. Unlike $J_{\rm H}$, which promotes parallel alignment of spins in different orbitals, it is introduced with the opposite sign, favoring antiparallel alignment and thus antiferromagnetic coupling across the bilayer~\cite{lechermann2025}.

Filled with three electrons, the introduced model is then solved using the dual triply irreducible local expansion (\mbox{D-TRILEX}) method~\cite{PhysRevB.100.205115, PhysRevB.103.245123, 10.21468/SciPostPhys.13.2.036}.
This approach enables a consistent treatment of the leading non-local collective electronic fluctuations by performing a diagrammatic expansion beyond the DMFT starting point, which accounts for the local correlation effects numerically exact.
The key advantage of this method lies in its efficient diagrammatic structure, which incorporates self-consistently the effect of spatial fluctuations in the charge, spin, and orbital channels on the electronic spectral function~\cite{stepanov2021coexisting, Vandelli2024_PbSi, PhysRevB.110.L161106, PhysRevLett.132.236504, j6bj-gz7j, Cuprates, FM_doping}, and remains computationally feasible even in the multi-band~\cite{PhysRevLett.127.207205, PhysRevLett.129.096404, PhysRevResearch.5.L022016, PhysRevLett.132.226501, stepanov2024charge, ts6y-zb6m, dhm8-5ss6} and real-time~\cite{vglv-2rmv, gzwp-zd6t} frameworks.
The \mbox{D-TRILEX} calculations are performed using the implementation presented in Ref.~\cite{10.21468/SciPostPhys.13.2.036}. 
The DMFT impurity problem is solved using the continuous time quantum Monte Carlo~\cite{PhysRevB.72.035122, PhysRevLett.97.076405, PhysRevLett.104.146401, RevModPhys.83.349} solver implemented in \textsc{w2dynamics} package~\cite{wallerberger2019}.
All calculations are performed on a ${36\times36}$ ${\bf k}$-grid in the Brillouin zone (BZ).
The local and momentum-resolved electronic functions are obtained from the corresponding Matsubara Green's functions via analytical continuation using the maximum entropy method implemented in the \textsc{ana\_cont} package~\cite{kaufmann2021anacont}. 
The orbital-resolved charge (${\varsigma={\rm ch}}$) and spin (${\varsigma={\rm sp}}$) susceptibilities are calculated as:
$
X^{\varsigma}_{ll'}({\bf q},\omega) = \langle n^{\varsigma}_{{\bf q},\omega,l} \, n^{\varsigma}_{{\bf -q},-\omega,l'} \rangle-\langle n^{\varsigma}_{l}\rangle \langle n^{\varsigma}_{l'}\rangle, 
$
where ${n^{\rm ch/sp} = n_{\uparrow} \pm n_{\downarrow}}$ is the charge/spin density.

\begin{figure}[t!]
\includegraphics[width=1\linewidth]{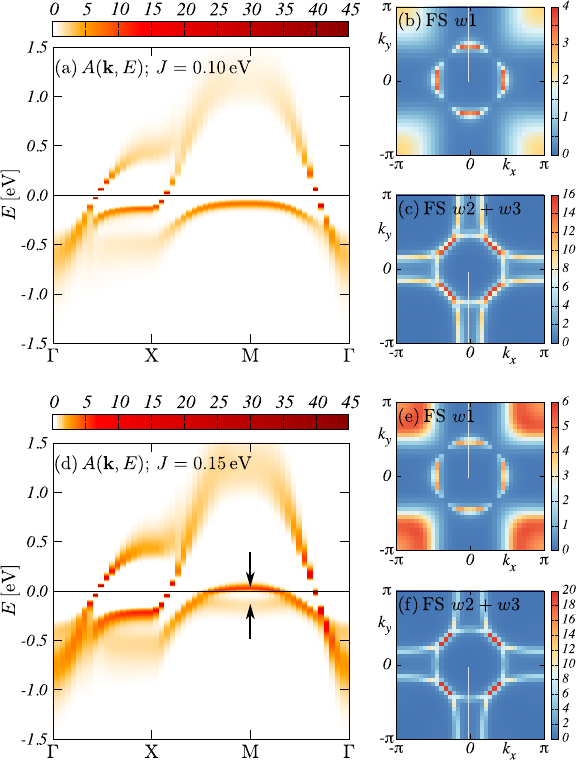}
\caption{Calculated momentum-resolved electronic spectral function $A({\bf q},E)$ (a,\,d) obtained along the high-symmetry path, and the imaginary part of the Green's function at the lowest Matsubara frequency [${-\text{Im}\,G({\bf k},\nu_0)}$, approximate FS] calculated for the $w1$ (b,\,e) and the sum of $w2$ and $w3$ (c,\,f) orbitals.
The results are obtained at at ${T=166}$\,K for ${J=0.10}$\,eV (a-c) and ${J=0.15}$\,eV (d-f). The black arrows in panel (d) highlight the splitting of the flat part of the $\gamma$ band and the formation of the shadow band below the Fermi level. 
\label{fig:spec_func}}
\end{figure}

{\it Results.}
To determine realistic values for the interaction parameters $U$ and $J$, we first perform calculations at ${T=290}$\,K.
Consistent with previous studies slave-boson  studies~\cite{lechermann2025}, we find that increasing $J$ reduces the occupancy of the $\gamma$ band, which shifts its flat part closer to the Fermi level. 
The appearance of this flat band near the Fermi energy leads to a charge ordering (CO) instability, which, for a fixed ratio ${U/J=40}$, occurs at a critical interaction strength of ${U_{\rm CO}\simeq{}6.5}$\,eV.
Increasing $U$ also renormalizes the bandwidth of the electronic dispersion. 
These two criteria fix the interaction strength at ${U = 5.5}$\,eV, giving a renormalization factor of about two for the electronic dispersion, consistent with realistic DMFT calculations~\cite{lechermann23,shilenko23}, while remaining below the threshold for the CO instability.

\begin{figure}[t!]
\includegraphics[width=1\linewidth]{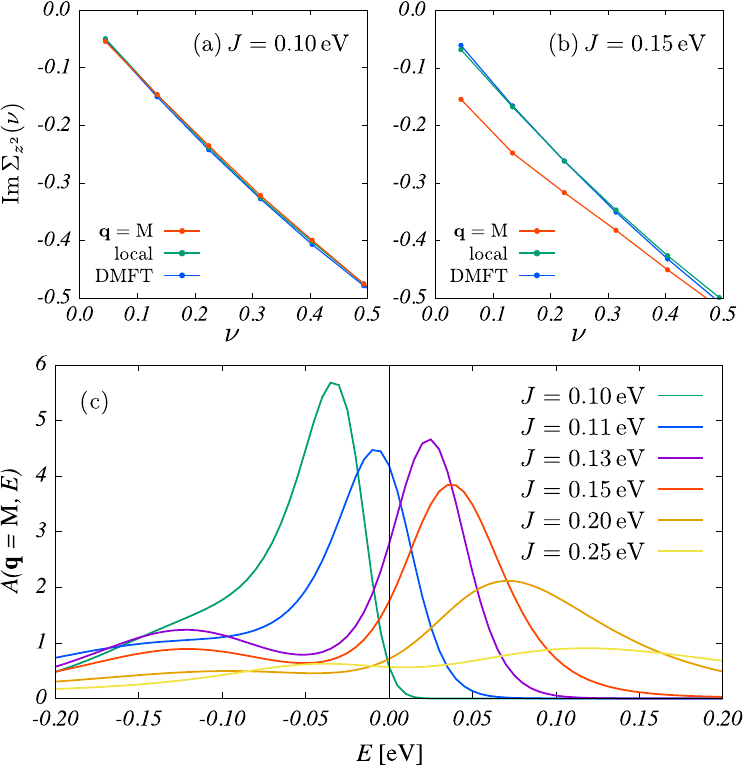}
\caption{(a, b) Calculated imaginary part of the electronic self-energy of the $w1$ orbital as a function of Matsubara frequency $\nu$. The results are obtained at ${T=166}$\,K for ${J=0.10}$\,eV (a) and ${J=0.15}$\,eV (b). The red curve corresponds to the ${{\bf q}=\text{M}}$ point, the green one depicts the local self-energy, and the blue curve corresponds to the DMFT solution.
(c) The electronic spectral function ${A({\bf k}=\text{M}, E)}$ of the $w1$ orbital obtained at ${T=290}$\,K for different $J$. 
At ${J=0.13}$\,eV and ${J=0.15}$\,eV the spectral function exhibits a double-peak structure, indicating the formation of a spin-polaron shadow band.
%\IE{Observe the double-peak structure (formation of the polaronic shadow band) for $J\geq 0.13$\,eV.}
\label{fig:M_point}}
\end{figure}

The tendency toward the CO state also increases upon lowering the temperature (see Supplemental Material (SM)~\cite{SM}). 
We find, that at ${U=5.5}$\,eV and ${J=U/40}$ the CO instability appears at ${T\simeq150}$\,K.
For this reason, the calculations are performed in the normal state above ${T=150}$\,K.
We keep the value of $J$ as a tuning parameter, that controls the occupation and hence the location of the $\gamma$ band. The latter is believed to be of vital importance for the formation of superconducting state~\cite{ncbf-9b8m}.
For a smaller value of ${J=0.10}$\,eV, the flat part of the $\gamma$ band, which appears in the vicinity of the ${\text{M}=(\pi,\pi)}$ point, lies close, but yet below the Fermi energy. 
The corresponding electronic spectral function, obtained at ${T=166}$\,K, is shown in Fig.~\ref{fig:spec_func}\,(a).
In this case, the orbital occupations read ${n_{1}=1.74}$ and ${n_{2}=n_{3}=0.63}$, while the spectral function likewise resembles certain available DMFT results, e.g.~\cite{ryee24}.
This similarity can be attributed to the fact that non-local charge and spin fluctuations are rather weak in this regime.
The strength of these fluctuations can be estimated by the leading eigenvalue (LE) of the Bethe-Salpeter equation, that accounts for the particle-hole fluctuations in the corresponding channel. 
The ${\text{LE}=0}$ indicates the absence of fluctuations, while ${\text{LE}=1}$ results in a divergence, associated with the spontaneous symmetry breaking, and signals the formation of the ordered state.
%with the wave vector ${\bf q}$ at which this divergence occurs.
At ${T\simeq166}$\,K we find that ${\text{LE}_{\rm ch}=0.29}$ and ${\text{LE}_{\rm sp}=0.36}$, which indicates that the non-local charge and spin fluctuations are indeed weak.
Consequently, these results a nearly momentum-independent \mbox{D-TRILEX} self-energy, similar to that of DMFT [see Fig.~\ref{fig:M_point}\,(a)].

In Fig.~\ref{fig:spec_func}\,(b,\,c) we plot the calculated imaginary part of the intraband Green's function obtained at the lowest Matsubara frequency ${\nu_0=\pi/\beta}$, ${-\text{Im}\,G_{ll}({\bf k},\nu_0)}$, which approximates the Fermi surface (FS) as projected onto the $w1$ (b) and the sum of the $w2$ and $w3$ (c) orbitals. 
We find, that the largest spectral weight of the $w1$ orbital at Fermi energy is located at the small circle around the $\Gamma$ point of the BZ ($\alpha$-sheet of the FS). A small spectral weight in the vicinity of M point originates from the $\gamma$ band part located in close proximity to the Fermi level.
The spectral weight of the $w2$ and $w3$ orbitals is distributed over the $\alpha$ sheet and the cross-shaped structure corresponding to the $\beta$ sheet of the FS.

\begin{figure}[t!]
\includegraphics[width=1\linewidth]{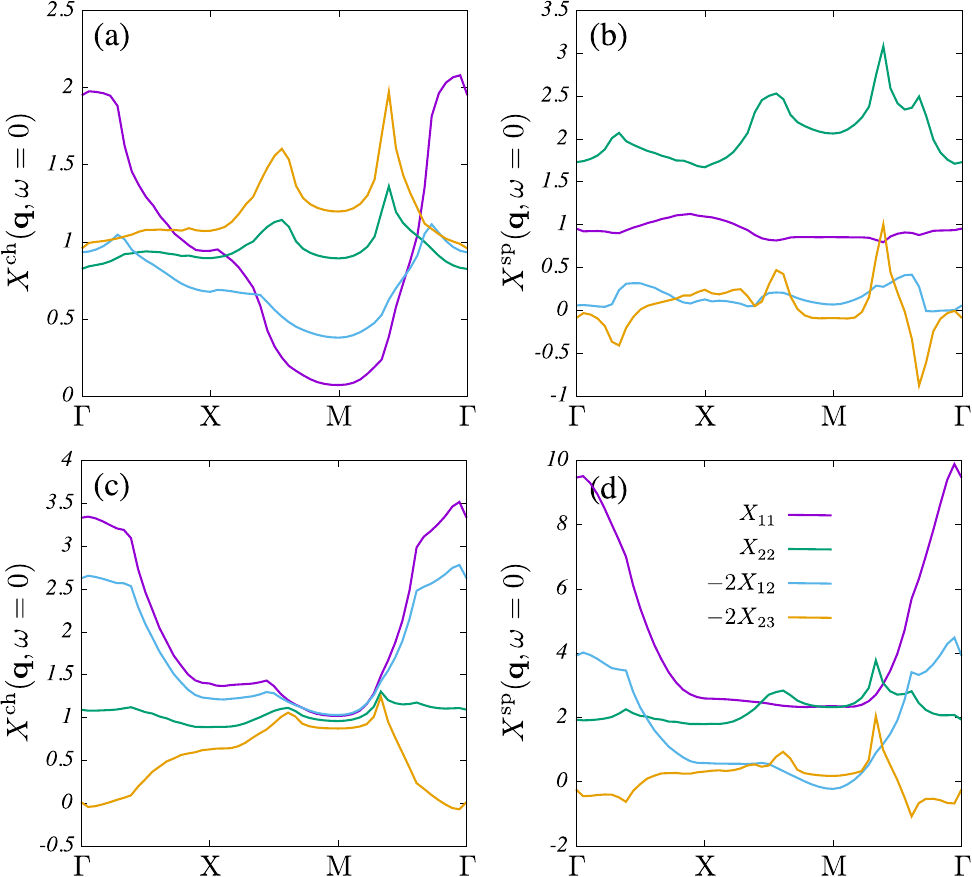}
\caption{Calculated momentum- and orbital-resolved static (${\omega=0}$) charge (a,\,c) and spin (b,\,d) susceptibilities.
The results are obtained at ${T=166}$\,K for ${J=0.10}$\,eV (a,\,b) and ${J=0.15}$\,eV (c,\,d) along the high-symmetry path in momentum space. The interorbital susceptibilities are plotted with the negative sign. 
\label{fig:X_tot}}
\end{figure}

This form of the FS explains the distinct character of charge and spin fluctuations originating from different orbitals.
The corresponding static orbital-resolved charge (a) and spin (b) susceptibilities $X^{\varsigma}_{ll'}({\bf q},\omega=0)$ are shown in Fig.~\ref{fig:X_tot}.
We find that the charge fluctuations associated with the $w1$ orbital correspond to small momenta, with the susceptibility showing its maximum near the ${\Gamma=(0,0)}$ point (magenta curve). 
This behavior is linked to the nesting within the $\gamma$ sheet of the FS located in the vicinity of the M point [Fig.~\ref{fig:spec_func}\,(b)].
In turn, the charge and spin fluctuations originating from the remaining $w2$ and $w3$ orbitals are incommensurate, with the susceptibility exhibiting a maximum at the wave vector ${{\bf q}=(\frac{2\pi}{3},\frac{2\pi}{3})}$, corresponding to the nesting between regions of largest spectral weight in the $\alpha$ sheet [Fig.~\ref{fig:spec_func}\,(c)].
Remarkably, we observe that the inter- and intraorbital charge fluctuations have opposite signs but nearly identical momentum dependence. 
Indeed, $-X^{\rm ch}_{12}$ and $-X^{\rm ch}_{23}$ closely resemble $X^{\rm ch}_{11}$ and $X^{\rm ch}_{22}$, respectively.
In contrast, the spin fluctuations are dominated by intraorbital scattering processes. To physically interpret the peaks in the susceptibility, one should note that $X_{22} \pm X_{23}$ contributes to the symmetric (even) and antisymmetric (odd) channel \cite{botzel2024,SM}.  
We also find that for ${J=0.10}$\,eV the leading charge and spin fluctuations originate from the $w2$ and $w3$ orbitals. 
The fluctuations, especially the magnetic ones, related to the $w1$ band are suppressed because this orbital is nearly fully filled, and the flat part of the $\gamma$ dispersion carrying large spectral weight lies below the Fermi energy, thus contributing little to particle–hole scattering processes.

We now turn to the behavior of the system at larger value of $J$.
Increasing $J$ reduces the occupation of the $\gamma$ band (related to $n_1$) and shifts the flat part of this band upward in energy.
As shown in Fig.~\ref{fig:M_point}\,(c), at ${J=0.11}$\,eV (${n_{1}=1.71}$) the flat part of the $\gamma$ band lies close to the Fermi energy, while for ${J>0.12}$\,eV it shifts to positive energies.
At ${J=0.12}$\,eV, the flat band appears exactly at the Fermi level, and the system becomes unstable toward the formation of the CO state.
Remarkably, we observe that once the flat part of the $\gamma$ band crosses the Fermi level (at ${J=0.13}$\,eV, ${n_{1}=1.62}$), it immediately splits into two branches, the main peak above $E_F$ and the weaker (shadow) peak appearing below the Fermi energy [see Fig.~\ref{fig:M_point}\,(c)].
With further increase of $J$, the intensity of both branches progressively diminishes (${J=0.15}$\,eV, ${n_{1}=1.55}$). 
The lower branch vanishes at ${J=0.20}$\,eV (${n_{1}=1.36}$), while the upper branch becomes washed out at ${J=0.25}$\,eV (${n_{1}=1.22}$).

The full momentum-resolved electronic spectral function calculated at ${J=0.15}$\,eV and ${T=166}$\,K is shown in Fig.~\ref{fig:spec_func}\,(d).
We observe, that besides the splitting of the flat band, indicated by the black arrows, the rest of the spectral function remains nearly unchanged compared to the ${J=0.10}$\,eV case, shown in panel (a).
This behavior indicates that the splitting is driven by non-local (i.e., momentum-dependent) fluctuations, which is further supported by the form of the electronic self-energy.
At ${J=0.15}$\,eV, the self-energy of the $w1$ orbital becomes strongly momentum-dependent and even exhibits non-Fermi-liquid behavior at the M point [see Fig.~\ref{fig:M_point}\,(b)].
On the contrary, while the self-energy of the remaining $w2$ and $w3$ is also momentum-dependent, it coincides with the local DMFT result at ${\bf k}$-points corresponding to the FS (see SM~\cite{SM}). 

We attribute the emergence of momentum dependence in the self-energy with the enhanced strength of magnetic fluctuations. 
Indeed, increasing $J$ from ${J=0.10}$\,eV to ${J=0.15}$\,eV only slightly increases the strength of the charge fluctuations, from ${\text{LE}_{\rm ch}=0.29}$ to ${\text{LE}_{\rm ch}=0.32}$. 
In contrast, the strength of spin fluctuations increases substantially, from ${\text{LE}_{\rm sp}=0.36}$ to ${\text{LE}_{\rm sp}=0.63}$.
These strong fluctuations originate from the $\gamma$ band, which now has a large spectral weight at the Fermi energy.
Indeed, we find that the largest contributions to the charge and spin susceptibilities come from intraband scattering processes from the $w1$ orbital and exhibit a FM-like character [Fig.~\ref{fig:X_tot}\,(c,\,d)].
This behavior can be attributed to nesting-like features within the $\gamma$ sheet, whose spectral weight is significantly enhanced upon increasing $J$ from ${J=0.10}$\,eV to ${J=0.15}$\,eV [Fig.~\ref{fig:spec_func}\,(e)].
The spectral weight of the $\alpha$ sheet, associated with the $w2$ and $w3$ orbitals, also increases, but the nested (red) area is noticeably reduced. This leads to a suppression of collective electronic fluctuations related to the $w2$ and $w3$ orbitals compared to those originating from the $w1$ orbital.
As in the case of ${J=0.10}$\,eV, the inter- and intraband charge susceptibilities have opposite signs but nearly identical momentum dependence, whereas the spin susceptibility remains dominated by intraband scattering.

This result suggests that the splitting of the flat part of the $\gamma$ band is primarily driven by low-{\bf q} FM-like spin fluctuations. 
To confirm this, in SM~\cite{SM} we plot the imaginary part of the electronic self-energy across the entire BZ at the lowest Matsubara frequency alongside the imaginary part of the Green's function and find that they exhibit very similar momentum dependence. 
This indicates that the main contribution to the self-energy arises from electron scattering on collective fluctuations with zero momentum transfer, which is described by the product of the Green's function $G({\bf k+Q})$ with ${{\bf Q}=(0,0)}$ and the renormalized spin-channel interaction at zero momentum, $W^{\rm sp}({\bf Q})$. 
These fluctuations induce non-Fermi-liquid-like behavior of the self-energy in the region of the M point, leading to the formation of an incoherent FM spin-polaron band (bound state) below the Fermi energy, similar to the mechanism discussed previously in other systems~\cite{katsnelson82, PhysRevB.91.155114, FM_doping}.
The position of the spin-polaron band changes only slightly with decreasing temperature (see SM~\cite{SM}).
Note that the splitting of the $\gamma$-based spectral function and the formation of the shadow-band peak below $E_F$ could potentially explain the conflicting results from angle-resolved photoemission spectroscopy (ARPES) measurements, which report different location of the $\gamma$ pocket to be either below or above the Fermi level~\cite{sun2025observationsuperconductivityinducedleadingedgegap,wang2025electronicstructurecompressivelystrained,Li2025}. An experimental test for the formation of the spin-polaron band could be the strong temperature-dependent increase of the uniform spin susceptibility, showing a tendency towards FM like order. Optics measurements may also prove helpful to detect this composite excitations~\cite{wangzheng04}. In addition, the formation of the polaronic band is expected to occur upon hole doping, i.e. once the $\gamma$ pocket crosses the Fermi level.

{\it Conclusion.}
To conclude, we find that the charge and spin fluctuations in La$_3$Ni$_2$O$_7$ are strongly orbital- and momentum-dependent, and are highly sensitive to the position of the flat $\gamma$ band in the electronic spectrum. 
The latter can be tuned via the occupation of the $w1$ orbital, which in this work is controlled by varying the effective interorbital interaction $J$.
At smaller $J$, when the flat portion of the $\gamma$ band lies below the Fermi energy, the strongest collective fluctuations mainly originate from the $w2$ and $w3$ orbitals. 
Upon increasing $J$, the flat band approaches the Fermi energy and the system becomes unstable toward the formation of a CO state. 
Further increase of $J$ shifts the flat band above the Fermi level, thereby altering the balance of orbital contributions to the charge and spin fluctuations. 
The leading fluctuations then arise from the $\gamma$ band, and the dominant spin fluctuations become FM-like.
Electronic scattering on the latter results in a non-Fermi-liquid self-energy and this leads to the formation of a FM spin-polaron band below the Fermi energy.
This could potentially explain the controversy of the recent ARPES experiments regarding the position of the $\gamma$ pocket with respect to the Fermi level~\cite{sun2025observationsuperconductivityinducedleadingedgegap,wang2025electronicstructurecompressivelystrained,Li2025}. We further propose experimental tests to validate the spin-polaron formation in bilayer nickelate.

{\it Note added:} After completion of this work, we were informed about to-be-published ARPES data which reveals flat spectral weight appearing below the Fermi level upon shifting the $\gamma$ band across the Fermi level~\cite{PC}.

\begin{acknowledgments}
E.A.S. acknowledges the support from ANR JCJC grant ``ELECTRO'', ANR-25-CE30-7064.
\end{acknowledgments}

\bibliography{Ref}

\end{document}

\title{Supplemental Material\\[0.3cm]
Co-operating multiorbital and nonlocal correlations in bilayer nickelate} 

\author{Evgeny A. Stepanov}
%\email{evgeny.stepanov@polytechnique.edu}
\affiliation{CPHT, CNRS, \'Ecole polytechnique, Institut Polytechnique de Paris, 91120 Palaiseau, France}
\affiliation{Coll\`ege de France, 11 place Marcelin Berthelot, 75005 Paris, France}
\author{Steffen B\"otzel}
\affiliation{Theoretische Physik III, Ruhr-Universit\"at Bochum, D-44780 Bochum, Germany}
\author{Ilya M. Eremin}
\affiliation{Theoretische Physik III, Ruhr-Universit\"at Bochum, D-44780 Bochum, Germany}
  \author{Frank Lechermann}
\affiliation{Theoretische Physik III, Ruhr-Universit\"at Bochum,
  D-44780 Bochum, Germany}  

\begin{abstract}
~
\end{abstract}

\maketitle
\twocolumngrid

\begin{figure}[t!]
\includegraphics[width=1\linewidth]{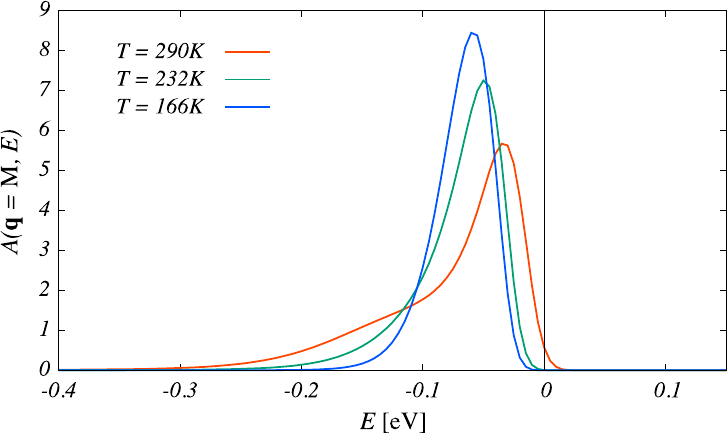}
\caption{Calculated electronic spectral function ${A({\bf k}=\text{M}, E)}$ obtained for the $w1$ band at ${T=290}$\,K (red), ${T=232}$\,K (green), and ${T=166}$\,K (blue) for the case of ${J=0.10}$\,eV.
\label{fig:M_J0.10}}
\end{figure}

\section{Temperature dependence of the flat $\gamma$ band}

\begin{figure}[t!]
\includegraphics[width=1\linewidth]{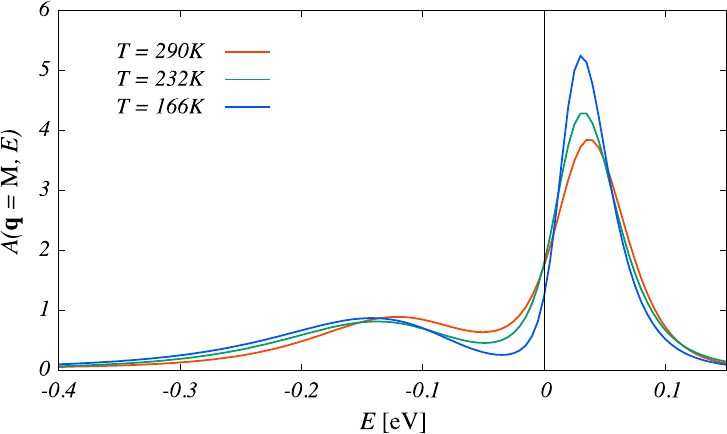}
\caption{Calculated electronic spectral function ${A({\bf k}=\text{M}, E)}$ obtained for the $w1$ band at ${T=290}$\,K (red), ${T=232}$\,K (green), and ${T=166}$\,K (blue) for the case of ${J=0.15}$\,eV.
\label{fig:M_J0.15}}
\end{figure}

In this section we examine the temperature dependence of the flat $\gamma$ band. 
To this end, we analyze the electronic spectral function of the $w1$-orbital at the ${{\bf k}=\text{M}}$ point for three temperatures, ${T=290}$\,K (red), ${T=232}$\,K (green), and ${T=166}$\,K (blue).
The corresponding results for ${J=0.10}$\,eV and ${J=0.15}$\,eV are shown in Figs.~\ref{fig:M_J0.10} and~\ref{fig:M_J0.15}, respectively.
In both cases, the spectral weight of the flat band peak, i.e. the main peak, appearing below (for ${J=0.10}$\,eV) or above (for ${J=0.15}$\,eV) the Fermi energy, increases upon decreasing temperature.
Consequently, this enhances the particle-hole fluctuations in the charge channel end eventually leads to the formation of the CO state below ${T\approx150}$\,K, as discussed in the main text.
Remarkably, while at ${J=0.10}$\,eV the flat band shifts slightly to the lower energies upon decreasing temperature, at ${J=0.15}$\,eV the positions of both the flat band and its spin-polaron satellite (shadow band) remain essentially unchanged. 
Moreover, the intensity of the spin-polaron band exhibits negligible temperature dependence.

\section{Identification of the leading scattering momentum of magnetic fluctuations}

\begin{figure}[t!]
\includegraphics[width=1\linewidth]{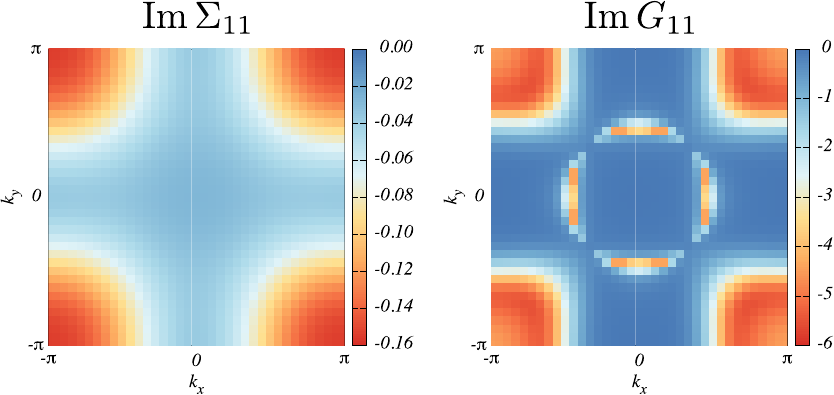}
\caption{Calculated imaginary part of the electronic self-energy (left panel) and of the Green's function (right panel) from the $w1$ band.
The result is obtained for ${J=0.15}$, ${T=145}$\,K at the lowest Matsubara frequency $\nu_0$. 
\label{fig:SG}}
\end{figure}

In the main text, we identified the spin-polaron band arising from the electronic scattering on the ferromagnetic (FM) fluctuation associated with the $w1$-orbital. 
To confirm this mechanism, in Fig.~\ref{fig:SG} we compare the calculated momentum dependence of the imaginary part of the intra-band $w1$ self-energy with that of the corresponding Green's function, evaluated at the lowest Matsubara frequency across the first Brillouin zone. 
Within a $GW$-like approximation, the dominant contribution of magnetic fluctuations to the self-energy reads:
\begin{align}
\Sigma({\bf k},\nu) \simeq
W^{\rm sp}({\bf Q}), G({\bf k+Q},\nu),
\end{align}
describing electron scattering, encoded in the Green's function $G$, on the leading magnetic fluctuation (the renormlalized interaction $W^{\rm sp}$ in the spin channel, proportional to the spin susceptibility) at wave vector ${\bf Q}$. 
We find that both quantities exhibit nearly identical momentum dependence, except that the self-energy does not reveal additional features at the $\alpha$ sheet of the Fermi surface (the small pocket around $\Gamma$-point, visible in $G$). 
This indicates that the dominant contribution to the self-energy originates from scattering on FM fluctuations with vanishing momentum transfer, ${{\bf Q}=0}$.

\begin{figure}[t!]
\includegraphics[width=1\linewidth]{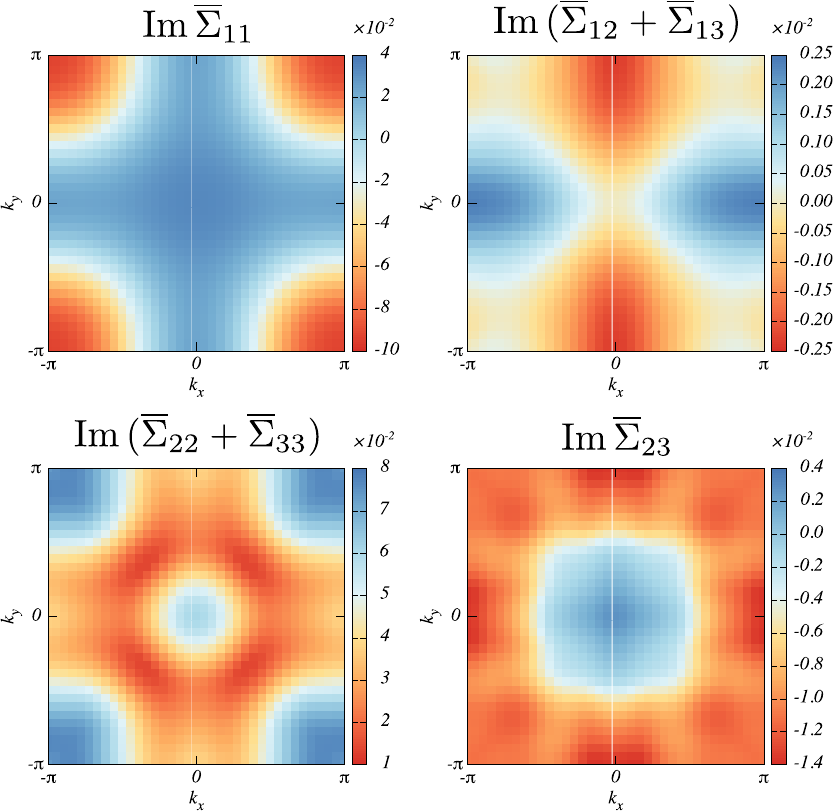}
\caption{Calculated difference between the imaginary parts of the \mbox{D-TRILEX} and DMFT self-energies. The result is obtained at the lowest Matsubara frequency $\nu_0$ for different orbitals. The calculations are performed at ${T=166}$\,K for ${J=0.15}$\,eV. 
\label{fig:S_BZ_orb}}
\end{figure}

\section{Orbital-and momentum-resolved self-energy}

In this section, we analyze the orbital and momentum dependence of the electronic self-energy. 
For ${J=0.10}$\,eV, the \mbox{D-TRILEX} results closely resemble those of DMFT. 
We therefore focus on ${J=0.15}$\,eV, where significant deviations from DMFT arise. 
In Fig.~\ref{fig:S_BZ_orb}, we show the difference between the imaginary parts of the non-local \mbox{D-TRILEX} and local DMFT self-energies for different orbital components. 
We find, that the inter-orbital components are much smaller than the intra-orbital ones. 
The latter exhibit the strongest momentum variation near the M point, where the $w1$ band lies close to the Fermi energy. 
This negative (relative to DMFT) variation leads to non-Fermi-liquid behavior and the emergence of the spin-polaron band, as discussed in the main text. 
In contrast, the $w2$ and $w3$ bands are far from the Fermi energy at the M point, and at momenta corresponding to their Fermi surfaces the \mbox{D-TRILEX}–DMFT difference remains negligible. 
These results demonstrate that electronic correlations become strongly orbital and momentum dependent when the flat part of the $w1$ band crosses the Fermi level.
\\\vspace{8cm}~